\documentclass[prl,aps,twocolumn]{revtex4}
\usepackage{graphicx} 
\usepackage[usenames]{color}
\usepackage{amsmath,amssymb}
\usepackage{natbib}
\usepackage{color}
 \usepackage[normalem]{ulem}
 \usepackage[geometry]{ifsym} 
\def\strutdepth{\dp\strutbox}
\def\nw#1{\strut\vadjust{\kern-\strutdepth\vtop to0pt{\vss\hbox to\hsize
{\hskip\hsize\hskip5pt$\leftarrow$\hss\strut}}}{\em #1}}

\begin{document}
\title{Initial spreading of low-viscosity drops on partially wetting surfaces}
\author{Koen G. Winkels, Joost H. Weijs, Antonin Eddi, and Jacco H. Snoeijer}
\affiliation{Physics of Fluids Group, Faculty of Science and Technology and Mesa+ Institute, University of Twente, 7500 AE Enschede, The Netherlands}
\date{\today}

\begin{abstract}
Liquid drops start spreading directly after brought into contact with a partial wetting substrate. Although this phenomenon involves a three-phase contact line, the spreading motion is very fast. We study the initial spreading dynamics of low-viscosity drops, using two complementary methods: Molecular Dynamics simulations and high-speed imaging. We access previously unexplored length- and time-scales, and provide a detailed picture on how the initial contact between the liquid drop and the solid is established. Both methods unambiguously point towards a spreading regime that is independent of wettability, with the contact radius growing as the square root of time.

\end{abstract}

\maketitle
How fast can a liquid drop spread over a surface? This basic question is relevant for applications ranging from printing and coating, to agricultural applications \cite{Wijshoff:2010, Simpkins:2003, Bonn:2009,Vovelle:2000, deGennes:1985}. In the final stage of drop spreading the dynamics are governed by Tanner's law, which relates the radius of the wetted area with time as $r \sim t^{1/10}$~\cite{Tanner:1979, Bonn:2002}. This extremely slow dynamics emerges from a balance between surface tension and viscous forces close to the contact line \cite{Bonn:2009}. Much less is known about the early stages of spreading, just after a spherical drop is brought into contact with a substrate at vanishing approach velocity. In contrast to Tanner's law, this dynamics is very fast~\cite{Biance:2004,Bird:2008,Courbin:2009,Carlson:arXive,Carlson:2011}: capillary energy suddenly becomes available when the drop touches the solid, and this energy is concentrated into a singular point of contact. It has remained unclear whether or not the wetting conditions can influence such rapid inertial flows~\cite{Duez:2007,Eggers:2007, Duez:2010}.

The initial stages of drop spreading are strongly reminiscent of the coalescence of two spherical drops of liquid, which very rapidly merge after contact is established \cite{Eggers:1999, Eggers:2003,Wu:2004, Thoroddsen:2005, Case:2008, Paulsen:2011}. For low-viscosity liquids such as water, it is well-known that the contact area between the drops grows as $r \sim t^{1/2}$ during coalescence. This can be explained from the balance of the inertial pressure inside the drop, $\sim \rho (dr/dt)^2$, and the capillary pressure, $\sim \gamma R/r^2$. Here $\rho$ is the density, $\gamma$ the surface tension, and $R$ the drop radius. Interestingly, an identical scaling law was observed experimentally for water drops spreading on a \emph{completely wetting} surface~\cite{Biance:2004}; apparently, the presence of a three-phase contact line does not affect the pressure balance during the initial phase of spreading. A rather different picture emerged, however, for drops spreading on \emph{partially wetting} surfaces~\citep{Bird:2008,Courbin:2009,Carlson:2011}. The dynamics was found to depend strongly on surface wettability, $r \sim t^\alpha$, with a non-universal exponent $\alpha$ that varies with the equilibrium contact angle~\citep{Bird:2008}. This raises a number of intriguing questions: How can the contact line, and the surface chemistry, affect the ``coalescence'' of a drop with a surface? Are the initial stages of spreading truly non-universal, or is there a hidden regime at smaller times? How is contact established on a molecular scale? 

\begin{figure}[h]
	 \includegraphics[width=0.5\textwidth]{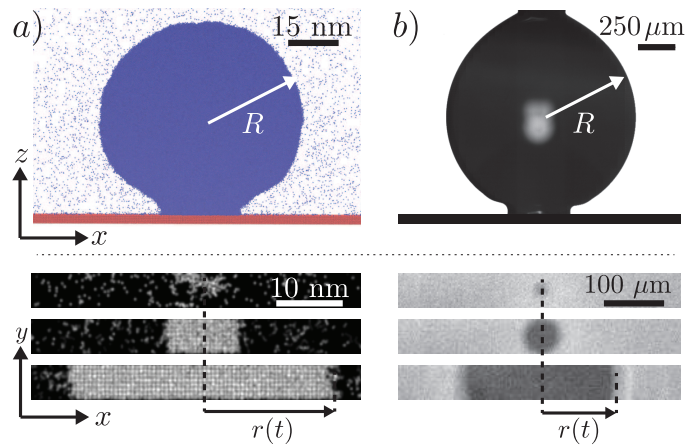}
	 \caption{\label{example} (Color online). Initial stages of drop spreading on partially wetting surfaces of varying wettability. (a) Molecular Dynamics simulations of Lennard-Jones nanodrops ($R=30$nm), and (b) Experiments of water drops ($R=0.5$mm). The top panels show side views of the liquid drop just after it has made contact with the partially wetting substrate. Lower panels are bottom views at times (a) $t=10, 35,400$ ps, and (b) $t=4,8,44$ $\mu$s. The contact radius $r(t)$ can be measured in time.}

\end{figure}

In this Letter we reveal the initial spreading dynamics of low-viscosity drops, using two complementary methods: Molecular Dynamics simulations of Lennard-Jones nanodrops and high-speed imaging of experiments on millimeter-sized water drops [Fig.~\ref{example}]. We access previously unexplored length- and time-scales, and provide a detailed picture on how the initial contact between the liquid drop and the solid is established. While simulations and experiments describe different dynamical regimes, both methods unambiguously point towards a universal spreading regime independent of wettability, consistent with the inertia-capillary balance $r \sim t^{1/2}$. This contrasts the scenario proposed by~\cite{Bird:2008}: At very early times after contact, the spreading exponent is independent of wettability, for contact angles ranging from complete wetting to very hydrophobic.

\paragraph{Molecular Dynamics simulations.---}The use of Molecular Dynamics simulations (MD) allows for studying the initial contact between a liquid drop and a solid substrate down to molecular scale. To reveal the fundamental mechanism of contact and subsequent spreading, we use a generic Lennard-Jones liquid. The advantage of the molecular approach is that, unlike continuum modeling, no assumptions on the moving contact line singularity~\cite{Bonn:2009} are needed. In MD, the wetting characteristics are directly controlled by the solid-liquid interaction, which determines the equilibrium contact angle $\theta_{eq}$~\cite{Weijs:2011}. The challenge, however, is to achieve sufficiently large drop sizes to recover a hydrodynamic regime. We therefore study a quasi-two-dimensional geometry rather than axisymmetric drops, in which the system size in the $y$-direction is only 15 molecular sizes [cf. Fig.~\ref{example}(b)]. Indeed, contact problems such as coalescence are known to be essentially 2D phenomena and the same is expected here~\cite{Eggers:1999, Burton:2007}. 

We perform simulations on binary systems, in which two types of particles exist: fluid particles that can move around either in the gas or liquid phase, and solid particles which are frozen on an fcc-lattice and constitute the solid substrate \cite{Spoel:2005}. All particle interactions are defined by the Lennard-Jones potential:

\begin{equation}
\phi_{ij}(r) =4\epsilon_{ij} \left[ \left( \frac{\sigma_{ij}}{r}\right)^{12} - \left( \frac{\sigma_{ij}}{r}\right)^{6}\right]\;.
\label{LJeq}
\end{equation}

\noindent Here, $\epsilon_{ij}$ is the interaction strength between particles $i$ and $j$ and $\sigma_{ij}$ the characteristic size of the atoms. This size is the same for all interactions, $\sigma_{ij} = \sigma = 0.34$~nm. The potential function is truncated at $r_{c} = 5\sigma$ ($1.7$ nm) where $\phi_{ij}$ is practically zero. The mass of the atoms was set at 20~amu, and a timestep of 1.75~fs was found to be sufficient to accurately model these systems. The interaction strengths between the fluid atoms are $\epsilon_{ll}=1.2k_BT$, with $k_B$ the Boltzmann constant and $T$ the temperature. The simulations are done in the NVT-ensemble, where the temperature is held at $300$K using a thermostat, which is below the critical point for a Lennard-Jones fluid with the interaction strengths used. 
%simulation details
The fluid particles (amount: $N_l = 304,192$) are initially positioned on an fcc-lattice (shaped with a cubic outline) far from the substrate ($N_s = 78,300$), but are free to move around and relax towards an equilibrium drop shape. Periodic boundary conditions are present in the lateral directions. The dimensions of the quasi-2D system are $240$~nm, $5.1$~nm and $120$~nm in the $x$-, $y$- and $z$-directions respectively (Fig.~\ref{example}). The depth of the system is short enough to suppress the Rayleigh-Plateau instability and leads to an infinitely long cylindrical-cap shaped drop.

%macroscopics
The interaction strength between the solid and fluid defines the contact angle \cite{Weijs:2011}.
We considered four different wettabilities (thus, four different values of $\epsilon_{sl}$): $\epsilon_{sl}=(0.3,0.4,0.8,1.2)k_BT$, giving $\theta_{eq}=(115^\circ,100^\circ,60^\circ,0^\circ) \pm 10^\circ$. The liquid density was measured to be $\rho = 664$~kg/m$^3$, and the surface tension $\gamma$ was measured in a separate, planar system at $\gamma = 0.017$~J/m$^2$. The viscosity of the liquid was measured in another separate system (Poiseuille-geometry) to be  $\eta =3.64\cdot 10^{-4}$~kg/(m$\cdot$s). 

\begin{figure}[ht]
	 \includegraphics[width=0.45\textwidth]{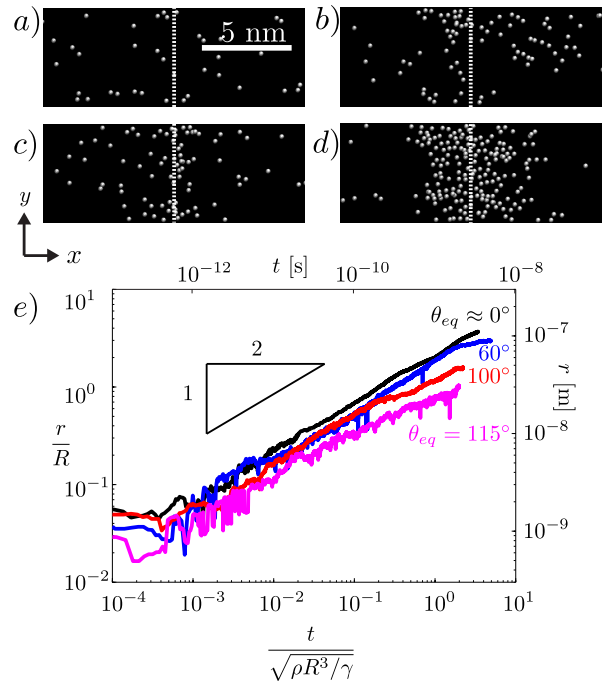}
	 \caption{\label{MDresults} (Color online). Molecular Dynamics results. (a-d) ``Bottom views" showing the molecules within $0.5$nm from the substrate just prior to and after initial contact ($t =-70$ ps, $-25$ ps, $-5$ ps and $35$ ps, respectively). (a) A small number of molecules from the vapor phase is close to the substrate. (b,c) Some fluctuating patches of higher density form. (d) A region of high liquid density nucleates at the substrate, from which one can measure $r(t)$. 
	 (e) Radius of wetted area as a function of time for varying substrate wettabilities $\theta_{eq}$. Once the contact is established, we observe a power-law with exponent 1/2 for all values of $\theta_{eq}$. Results are displayed in SI-units on the right and top axis, and displayed in dimensionless form on the left and bottom axis. Contact radius $r$ is rescaled with the initial drop radius $R$, time is rescaled with the inertial time scale $\tau_\rho = \sqrt{\rho R^3 /\gamma}$. }
\end{figure}

%physical procedure
The following procedure was used to bring the drop into contact with the substrate.
First, the liquid is allowed to equilibrate far away (32~nm) from the substrate.
During this stage, the drop will assume its cylindrical shape ($R = 30$~nm) and the liquid equilibrates with the vapour phase.
Next, a body force is briefly applied on the fluid atoms until the drop moves towards the substrate.
Just before the drop comes into contact with the substrate, the center-of-mass velocity of the drop is subtracted from the atom velocities such that the drop now 'hovers' above the substrate. Due to the close proximity of the substrate (around $1$~nm) the thermal fluctuations of the interface lead to first contact between the drop and the substrate after which the drop starts to spread.
Using this method, the approach velocity of the drop towards the substrate is zero and is not a parameter in this problem.

What happens during the initial contact? In these early stages one cannot yet speak about a continuous liquid phase in contact with the solid. Instead, one first encounters the discrete, molecular nature of the fluid. Figures~\ref{MDresults}(a-d) show snapshots of the molecules that are within $0.5$ nm from the substrate, represented as white dots. First a number of vapor molecules is randomly distributed over the surface [Fig.~\ref{MDresults}(a)]. As time progresses, more molecules come into contact and form fluctuating ``patches'' of high liquid density at the substrate [Figs.~\ref{MDresults}(b,c)]. The boundaries of these patches is extracted by computing the number density field of atoms near the surface, and taking the iso-density contour half-way between the liquid and vapor density. Eventually, the patch becomes sufficiently large to span the entire depth of the quasi-2D simulation domain, from which we define the time of contact [Fig. 2(d)]. The exact definition of $t=0$ does not influence our main conclusions below. From that moment, we track the boundaries of this wetting patch, which are the moving contact lines. The contact lines become sharper and are well-defined during the spreading, as can be seen in Fig.~\ref{example}(a). Note that while the average vapor density close to the surface is slightly larger than in the bulk, the surface coverage is very low and does not represent a precursor film.

The key result of our MD simulations is that, once the liquid drop has established contact with the surface, the spreading follows a single power law. Figure~\ref{MDresults}(e) shows the contact radius $r$ versus time on surfaces with varying wettability. For all contact angles $\theta_{eq}$ we observe a scaling  consistent with $r\sim t^{1/2}$ (best-fit exponent: $0.48$). It turns out that the contact angle of the substrate $\theta_{eq}$ does have an influence on the spreading, but only through the prefactor: the exponent is always very close to 1/2. The prefactor increases as the contact angle decreases, such that drops spread faster on the more hydrophilic surfaces. 

Despite the very small length- and time-scales in these simulations, the spreading appears to be consistent with the hydrodynamic picture of inertia-dominated coalescence~\cite{Eggers:1999,Wu:2004} The left axis and bottom axis in Fig.~\ref{MDresults}(e) represent the data in dimensionless units, where lengths are scaled with the initial drop radius $R$ and time with the inertial scale $\tau_\rho = \sqrt{\rho R^3 /\gamma}$. In these units, the data span a range similar to previous experiments on millimeter-sized water drops~\cite{Biance:2004,Bird:2008}. 

\begin{figure}[h]
	 \includegraphics[width=0.45\textwidth]{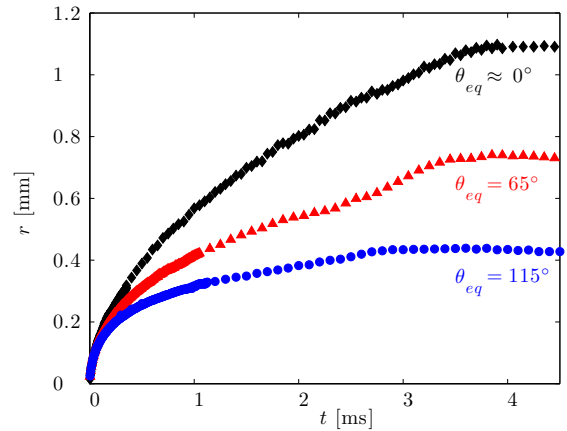}
	 \caption{\label{linlin} (Color online). Experimental measurements of contact radius $r$ plotted as function of time $t$ for different substrate wettabilities. Results for three different equilibrium contact angles are plotted: clean glass ($\theta_{e} \approx 0^{\circ}$; 
\textcolor{black}{\FilledDiamondshape}), coated glass ($\theta_{e} = 65 ^{\circ}$; \textcolor{red}{\FilledTriangleUp}), teflon coated glass ($\theta_{e} = 115^{\circ}$; \textcolor{blue}{\FilledCircle}). The curves represent averaged data of repeated measurements (five or more for each $\theta_{eq}$) per substrate for drops with radius $R=0.5$ mm, showing the reproducibility of the experiments. }
\end{figure}

\paragraph{Experiments.---}To verify whether the spreading behavior observed in MD is also found experimentally, we carried out experiments in a previously unexplored regime. The required spatial and temporal resolution is achieved by high-speed recording of drop spreading \emph{from below}, using transparent substrates and recording rates up to 600,000 frames/second. Typical images are shown in Fig.~\ref{example}b. The high-speed camera (Photron SA 1.1) is connected to a microscope (Zeiss Axiovert 25), which in combination with a 10X microscope objective (Zeiss A-plan, 10X) and reflective illumination gives a maximum resolving power of 2~$\mu$m/pixel. To capture a large period of the spreading process, frame rates used are in the range of 10-600kfps.
Qu\'{e}r\'{e} \emph{et al.} \cite{Biance:2004} and Bird \emph{et al.} \cite{Bird:2008}, have shown that data for different drop sizes collapse by inertial rescaling. As here we focus on the influence of wettability, we consider only one drop radius $R = 0.5 \pm 0.01$ mm.

To investigate the effect of wettability on the spreading, we performed experiments with water drops on three different substrates with different equilibrium contact angle $\theta_{eq}$: clean glass (almost perfectly wetting, $\theta_{eq}\approx 0^{\circ}$), coated glass ($\theta_{r} = 55^{\circ}$; $\theta_{a} = 75^{\circ}$; $\theta_{eq}=65^{\circ}$) and teflon coated glass ($\theta_{r} = 110^{\circ}$; $\theta_{a} = 120^{\circ}$; $\theta_{eq}=115^{\circ}$). In order to avoid any condensation effects prior to spreading, the surrounding air is saturated with nitrogen gas. A thin needle is fixed at height $D = 1$ $\pm$ $0.02$~mm above the substrate, thereby setting the initial radius $R$ of the spreading drop (height $D=2R$). With a syringe pump set at a constant volume rate of 1$\mu$l/min a pendant drop is grown at the needle tip, until it touches the substrate. This generates approach velocities $< 2 \cdot 10^{-5}$ m/s, so that the outer gas dynamics has a negligible influence on the contact process. The radius of contact $r(t)$ is determined from images as in Fig.~\ref{example}b, using a custom-made edge-detection algorithm in Matlab that finds the maximum image intensity slope in every frame.

\begin{figure}[h]
	 \includegraphics[width=0.45\textwidth]{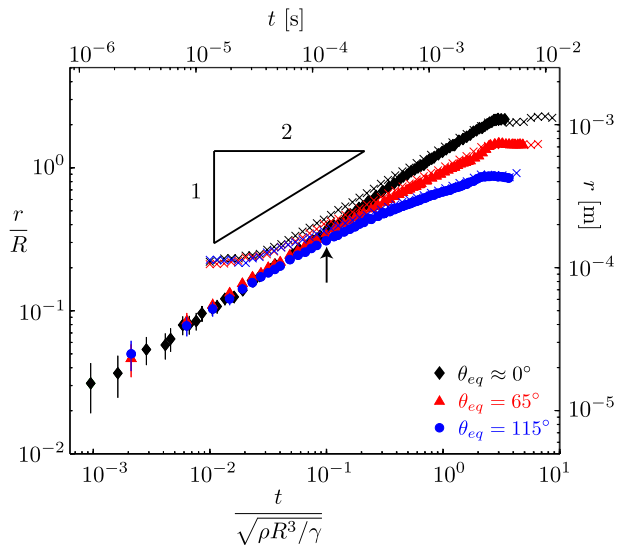}
	 \caption{\label{EXPresults} (Color online). Experimental results. Contact radius $r$ measured as a function of time (top and right axis) for three different equilibrium contact angles: clean glass ($\theta_{e} \approx 0^{\circ}$; 
\textcolor{black}{\FilledDiamondshape}), coated glass ($\theta_{e} = 65^{\circ}$; \textcolor{red}{\FilledTriangleUp}), and teflon coated glass ($\theta_{e} = 115^{\circ}$; \textcolor{blue}{\FilledCircle}). The data shown, is an average of at least 5 measurements. The error bars denote the statistical error, which is larger than the measurement accuracy. On the left and bottom axis the data is normalized by the drop radius $R$ and inertial time $\tau_\rho = \sqrt{\rho R^3 /\gamma}$ respectively. A new regime is observed at earlier times, where the spreading is \textit{independent} of the equilibrium contact angle. 
The colored crosses are reprinted data by Bird et al.~\cite{Bird:2008} (corresponding with: $\theta_{eq} =$ $3^{\circ}$, \textcolor{black}{$\times$}; $43^{\circ}$, \textcolor{red}{$\times$}; and $117^{\circ}$, \textcolor{blue}{$\times$}). The arrow indicates the smallest times that were accurately resolved in the study by Bird~\emph{et al.}}
\end{figure}

Our experiments confirm a single power law during the initial stages of contact. The measurements of  the contact radius $r(t)$ are shown in Fig.~\ref{linlin} on linear axes. One observes that the data fall onto three different curves, corresponding to the three values of $\theta_{eq}$. The curves separate about $0.1$~ms after contact, showing a dependence on wettability at later times. However, the early-time dynamics are independent of wettability. This is revealed in Fig.~\ref{EXPresults}, which shows the same data on log-log scale. We find that our data for different $\theta_{eq}$ perfectly collapse at early times ($t/\tau_\rho < 0.04$), and display an exponent close to $1/2$ (best fit: $0.55$). We reprinted the data by Bird~\emph{et al.}~\cite{Bird:2008} (crosses) for completeness, and find a perfect agreement with our data at $t/\tau_\rho > 0.1$, which is the range of accurate resolution in ref.~\onlinecite{Bird:2008}. 
The upper/right axis represent SI-units, while for the lower/left axis we employ the inertial scaling. Thus, the key point is that our measurements reveal a regime where wettability has no effect on the spreading at early times, not even in the prefactor.  
%%to be discussed
%Interestingly, neither the exponent nor the pre-factor is affected by the wettability. This implies that reaching the inertial regime after contact onwards, the dynamics are actually unique. Only at later stages, the inertial behaviour is affected in exponent, depending on the wetting characteristics of the substrate. 

\paragraph{Discussion.---} We have shown that early stage spreading of low-viscosity drops on a partially wetting substrate is independent of wettability. The wetted area is found to grow as $r \sim t^{1/2}$, for all considered wettabilities: we find no influence on the spreading exponent by the presence of a contact line. This suggests that the mechanism of capillary wave generation, invoked to explain $\theta_{eq}$-dependent spreading exponents~\cite{Bird:2008}, cannot be the dominant factor at very early times. Still, such capillary waves could be relevant for explaining the later stages of spreading in Fig.~\ref{EXPresults}, where a departure from the 1/2 scaling is observed. However, this departure first arises when $r/R \gtrsim 0.2$, in which case self-similarity of the bridge connecting the  liquid drop and the substrate is lost, and scaling cannot be assumed a priori. It therefore remains a challenge to explain the moment when the effect of $\theta_{eq}$ becomes apparent in the experiments.
There is, however, a subtle difference between the MD simulations and the experiments regarding the prefactor of the spreading law. For the experiments we observe a perfect collapse of the data on a single curve at early times. By contrast, the MD curves do not collapse, but the prefactor increases with decreasing $\theta_{eq}$. We can point out at least two possible origins for this difference. First, the time- and length-scales of the two systems differ by orders of magnitude. In addition the simulations and experiments are not dynamically similar. While the rescaled results of Fig.~\ref{MDresults} and~\ref{EXPresults} are very close, the Reynolds numbers defined as ${\rm Re} = \rho r (dr/dt)/\eta$ are very different: it is order unity in MD and order 100 in experiments. This suggests that the MD could be influenced by viscous effects, and it would be interesting to further investigate spreading for highly viscous liquids~\cite{Carlson:arXive,Paulsen:2011}. Another key difference is the importance of thermal fluctuations at molecular scales. These are known to have a dramatic effect on the dynamics of drop pinch-off~\cite{Moseler:2000,Eggers:2002}, and it would be interesting to further explore their influence on spreading in the molecular simulations. 

Finally, although wettability does not affect the initial rapid inertial flow in drop spreading, other cases are known to be strongly influenced by surface properties \cite{Duez:2007, Duez:2010, Tsai:2010}. From a more general perspective, the combination of such inertial flows with a three-phase contact line therefore remains a challenge. 

% If you have acknowledgments, this puts in the proper section head.

\begin{acknowledgments}
We gratefully acknowledge D. Lohse for discussions. R. van der Veen, M. van de Raa and J. Brons are thanked for their help during preliminary experiments. We are indebted to  J. Bird, S. Mandre, and H.A. Stone who granted permission to reprint their data.
This work is sponsored by the Foundation for Fundamental Research on Matter (FOM), National Computing Facilities Foundation (NCF) and VIDI-grant, which are financially supported by the Netherlands Organisation for Scientific Research (NWO). 
\end{acknowledgments}

% Create the reference section using BibTeX:


\begin{thebibliography}{10}

\bibitem{Wijshoff:2010}
H. Wijshoff, Physics Reports {\bf 491},  77  (2010).

\bibitem{Simpkins:2003}
P. Simpkins and V. Kuck, J. Coll. Int. Sci. {\bf 263},    (2003).

\bibitem{Bonn:2009}
D. Bonn, J. Eggers, J. Indekeu, J. Meunier, and E. Rolley, Rev. Mod. Phys. {\bf 81},  739  (2009).

\bibitem{Vovelle:2000}
V. Bergeron, D. Bonn, J.~Y. Martin, and L. Vovelle, Nature {\bf 405},
  (2000).

\bibitem{deGennes:1985}
P.~G. de~Gennes, Rev. Mod. Phys. {\bf 57},  827  (1985).

\bibitem{Tanner:1979}
L. Tanner, J. Phys. D: Appl. Phys. {\bf 12},  1  (1979).

\bibitem{Bonn:2002}
S. Rafa\"i, D. Sarker, V. Bergeron, J. Meunier, and D. Bonn, Langmuir {\bf 18},  10486  (2002).

\bibitem{Biance:2004}
A.-L. Biance, C. Clanet, and D. Qu\'er\'e, Phys. Rev. E {\bf 69},  016301
  (2004).

\bibitem{Bird:2008}
J.~C. Bird, S. Mandre, and H.~A. Stone, Phys. Rev. Lett. {\bf 100},  234501  (2008).

\bibitem{Courbin:2009}
L. Courbin, J.~C. Bird, M. Reyssat, and H.~A. Stone, J. Phys.: Condens. Matter
  {\bf 21},  464127  (2009).

\bibitem{Carlson:2011}
A. Carlson, M. Do-Quang, and G. Amberg, J. Fluid Mech. {\bf 682},  213  (2011).

\bibitem{Carlson:arXive}
A. Carlson, G. Bellani, and G. Amberg, Fluid Dynamics, arXiv:1111.1214v1 (2011)

\bibitem{Duez:2007}
C. Duez, C. Ybert, C. Clanet, and L. Bocquet, Nat. Phys. {\bf 3},  180
  (2007).

\bibitem{Eggers:2007}
J. Eggers, Nat. Phys. {\bf 3},  145  (2007).

\bibitem{Duez:2010}
C. Duez, C. Ybert, C. Clanet, and L. Bocquet, Phys. Rev. Lett. {\bf 104},
  084503  (2010).

\bibitem{Eggers:1999}
J. Eggers, J.~R. Lister, and H.~A. Stone, J. Fluid Mech. {\bf 401},  293
  (1999).

\bibitem{Eggers:2003}
L. Duchemin, J. Eggers, and C. Josserand, J. Fluid Mech. {\bf 487},  167  (2003).

\bibitem{Wu:2004}
M. Wu, T. Cubaud, and C.-M. Ho, Phys. Fluids {\bf 16},  L51  (2004).

\bibitem{Thoroddsen:2005}
S.~T. Thoroddsen, K. Takehara, and T.~G. Etoh, J. Fluid Mech. {\bf 527},  85
  (2005).

\bibitem{Case:2008}
S.~C. Case and S.~R. Nagel, Phys. Rev. Lett. {\bf 100},  084503  (2008).

\bibitem{Paulsen:2011}
J.~D. Paulsen, J.~C. Burton, and S.~R. Nagel, Phys. Rev. Lett. {\bf 106},
  114501  (2011).

\bibitem{Weijs:2011}
J.~H. Weijs, A. Marchand, B. Andreotti, D. Lohse, and  J.~H. Snoeijer, Phys. Fluids {\bf 23},  1  (2011).

\bibitem{Burton:2007}
J.~C. Burton and P. Taborek, Phys. Rev. Lett. {\bf 98},  224502  (2007).

\bibitem{Spoel:2005}
D. Van der Spoel, E. Lindahl, B. Hess, G. Groenhof, A.  Mark, and H. Berendsen, J. Comput. Chem. {\bf 26},  1701‚Äì1718  (2005).

\bibitem{Moseler:2000}
M. Moseler and U. Landman, Science {\bf 289},  1165  (2000).

\bibitem{Eggers:2002}
J. Eggers, Phys. Rev. Lett. {\bf 89},  084502  (2002).

\bibitem{Tsai:2010}
P. Tsai, R. C.~A. van~der Veen, M. van~de Raa, and D. Lohse, Langmuir {\bf 26},
   16090  (2010).

\end{thebibliography}
\end{document}